\pdfoutput=1

\documentclass[11pt,twoside,a4paper,cmspaper,final,collab]{cms-tdr}
\def\svnVersion{467385P}\def\cmsCernNoTag{CERN-EP-2017-186}\def\cmsCernDate{\today}\def\cmsMessage{Published in Physics Letters B as \href{http://dx.doi.org/10.1016/j.physletb.2018.05.074}{\doi{10.1016/j.physletb.2018.05.074.}}}

\begin{document}\cmsNoteHeader{HIN-16-001}

\hyphenation{had-ron-i-za-tion}
\hyphenation{cal-or-i-me-ter}
\hyphenation{de-vices}
\RCS$Revision: 456404 $
\RCS$HeadURL: svn+ssh://svn.cern.ch/reps/tdr2/papers/HIN-16-001/trunk/HIN-16-001.tex $
\RCS$Id: HIN-16-001.tex 456404 2018-04-19 20:50:17Z alverson $
\newlength\cmsFigWidth
\ifthenelse{\boolean{cms@external}}{\setlength\cmsFigWidth{0.85\columnwidth}}{\setlength\cmsFigWidth{0.4\textwidth}}
\ifthenelse{\boolean{cms@external}}{\providecommand{\cmsLeft}{top\xspace}}{\providecommand{\cmsLeft}{left\xspace}}
\ifthenelse{\boolean{cms@external}}{\providecommand{\cmsRight}{bottom\xspace}}{\providecommand{\cmsRight}{right\xspace}}

\newcommand{\sqrts}{\ensuremath{\sqrt{s}}\xspace}
\newcommand{\Dstar}{\ensuremath{\cmsSymbolFace{D}^{*+}}\xspace}
\newcommand{\PbPb}{\ensuremath{\mathrm{PbPb}}\xspace}
\newcommand{\pp}{\ensuremath{\Pp\Pp}\xspace}
\newcommand{\pPb}{\ensuremath{\Pp\mathrm{Pb}}\xspace}
\newcommand{\raa}{\ensuremath{R_{\mathrm{AA}}}\xspace}
\newcommand{\rpa}{\ensuremath{R_{\Pp\mathrm{A}}}\xspace}
\newcommand{\taa}{\ensuremath{T_{\mathrm{AA}}}\xspace}
\newcommand{\absy}{\ensuremath{ \abs{y}}\xspace}
\newcommand{\abseta}{\ensuremath{ \abs{\eta}}\xspace}
\newcommand{\Bplusminus}{\ensuremath{\cmsSymbolFace{B}^{\pm}}\xspace}

\cmsNoteHeader{HIN-16-001}
\title{Nuclear modification factor of $\PDz$ mesons in \PbPb collisions at $\sqrtsNN = 5.02\TeV$}

\date{\today}

\abstract{The transverse
momentum (\pt) spectrum of prompt \PDz{} mesons and their antiparticles has been measured
via the hadronic decay channels $ \PDz\to \PKm \PGpp$ and $\PADz \to \PKp \PGpm$ in \pp and \PbPb collisions at a centre-of-mass energy of 5.02\TeV per nucleon pair with the CMS detector at the LHC. The measurement is performed in the \PDz{} meson \pt range of 2--100\GeVc and in the rapidity range of $\absy<1$. The \pp (\PbPb{}) dataset used for this analysis corresponds to an integrated luminosity of 27.4\pbinv (530\mubinv). The measured \PDz{} meson \pt spectrum in pp collisions is well described by perturbative QCD calculations. The nuclear modification factor, comparing \PDz{} meson yields in \PbPb and \pp collisions, was extracted for both minimum-bias and the 10\% most central \PbPb interactions. For central events, the \PDz{} meson yield in the \PbPb collisions is suppressed by a factor of 5--6 compared to the \pp reference in the \pt range of 6--10\GeVc. For \PDz{} mesons in the high-\pt range of 60--100\GeVc, a significantly smaller suppression is observed. The results are also compared to theoretical calculations.
}
\hypersetup{%
pdfauthor={CMS Collaboration},%
pdftitle={Nuclear modification factor of D0 mesons in PbPb collisions at sqrt(s[NN]) = 5.02 TeV},%
pdfsubject={CMS},%
pdfkeywords={physics, suppression, quark gluon plasma, shadowing, D-meson, open heavy-flavour}
}

\maketitle

\section{Introduction}
Relativistic heavy ion collisions allow the study of quantum chromodynamics (QCD) at high energy density and temperature.
Lattice QCD calculations predict that under such extreme conditions a transition to a strongly interacting and deconfined medium, called the quark-gluon plasma (QGP), occurs~\cite{QGP1,QGP2,Karsch:2003jg}.
Heavy quarks are effective probes to study the properties of the deconfined medium created in heavy ion collisions.
These quarks are mostly produced in primary hard QCD scatterings with a production timescale that is shorter than the formation time of the QGP~\cite{PhysRevC.89.034906}. During their propagation through the medium, heavy quarks lose energy via radiative and collisional interactions with the medium constituents. Quarks are expected to lose less energy than gluons as a consequence of their smaller colour factor. In addition, the so-called ``dead-cone effect" is expected to reduce small-angle gluon radiation of heavy quarks when compared to both gluons and light quarks~\cite{Dokshitzer:2001zm,Armesto:2003jh,Andronic:2015wma}. Energy loss can be studied using the nuclear modification factor (\raa), defined as the ratio of the \PbPb yield to the \pp cross-section scaled by the nuclear overlap function~\cite{Miller:2007ri}.
Precise measurements of the \raa of particles containing both light and heavy quarks can thus provide important tests of QCD predictions at extreme densities and temperatures and in particular allow one to test the expected flavour dependence of the energy loss processes.
The comparison to theoretical calculations is fundamental in order to claim any evidence of flavour dependence of the energy loss mechanisms
since sizeable discrepancies in the \raa of light and heavy particles can arise as a consequence of the different transverse momentum spectra and fragmentation functions of beauty, charm, and light quarks and gluons.

Evidence of open charm suppression at the CERN LHC was observed by the ALICE Collaboration using the \raa
of promptly produced D mesons (\PDz, \PDp, \Dstar mesons and their conjugates) at mid-rapidity ($\absy<0.5$) at a nucleon-nucleon centre-of-mass energy $\sqrtsNN=2.76$\TeV. The measurement was performed as a function of centrality (\ie the degree of overlap of the two colliding nuclei) and transverse momentum ($1<\pt<36$\GeVc)~\cite{Adam:2015nna, Adam:2015sza}.
A maximum suppression by a factor of 5--6 with respect to the \pp reference was observed for the 10\% most central collisions at \pt of about 10\GeVc. A suppression by a factor of about 3 was measured at the highest \pt range studied, from 25 to 35\GeVc. The D meson \raa was found to be consistent with that for all charged particles for \pt from 6 to 36\GeVc. For lower \pt, the D meson \raa was observed to be slightly higher than the charged-particle \raa, although still compatible within the uncertainties~\cite{Abelev:2014laa, Abelev:2012hxa}.
At RHIC, the \raa of $\PDz$ mesons for the 10\% most central AuAu collisions at $\sqrtsNN=200$\GeV was measured by the STAR Collaboration in the rapidity range of $\absy<1$~\cite{PhysRevLett.113.142301}. A suppression by a factor of 2--3 for \pt larger than 3\GeVc was seen. This suggests that a significant energy loss of charm quarks in the hot medium also occurs at RHIC energies.
A first indication of a sizeable difference in the \raa of B and D mesons was observed when comparing the ALICE D meson \raa with the nonprompt \PJGy meson (i.e. from b-hadron decays) \raa measurement performed by the CMS Collaboration in \PbPb collisions at the same energy and collision centrality~\cite{Khachatryan:2016ypw}.
The \raa of nonprompt \PJGy mesons in the \pt range 6.5--30\GeVc was indeed found to be significantly larger than the \raa of D mesons in the 8--16\GeVc \pt region for central events. The \PDz{} \pt range was chosen to give a similar median \pt value to that of the parent b hadrons decaying to \PJGy particles~\cite{Adam:2015nna}.
Several measurements were also performed to address the relevance of cold nuclear matter effects for the suppression observed for heavy-flavour particles. Indeed, these phenomena can affect the yield of such particles, independently of the presence of a deconfined partonic medium. For instance, modifications of the parton distribution functions (PDFs) in the nucleus with respect to nucleon PDFs~\cite{Eskola:2009uj,deFlorian:2003qf,Frankfurt:2011cs} could change the production rate of heavy-flavour particles. To evaluate the relevance of these effects, the production of prompt D mesons was measured in \pPb collisions at mid-rapidity at 5.02\TeV by the ALICE Collaboration~\cite{Abelev:2014hha}. The nuclear modification factor in \pPb collisions (\rpa) was found to be consistent within the 15--20\% uncertainties with unity for \pt from 2 to 24\GeVc. This suggests that the suppression of D mesons observed in \PbPb collisions cannot be explained in terms of initial-state effects but is mostly due to strong final-state effects induced by the QGP.  A similar conclusion was obtained from the study of the \rpa of B mesons in \pPb collisions at 5.02\TeV, where values consistent with unity within the uncertainties were found for \pt from 10 to 60\GeVc~\cite{Khachatryan:2015uja}.

In this Letter, the production of prompt \PDz{} mesons in \PbPb collisions at 5.02\TeV is measured for the first time up to a \pt of 100\GeVc, allowing one to study the properties of the in-medium energy loss in a new kinematic regime. The \PDz{} meson and its antiparticle are reconstructed in the central rapidity region ($\absy<1$) of the CMS detector via the hadronic decay channels $\PDz\to \PKm \PGpp$ and $\PADz \to \PKp \PGpm$. The production cross section and yields in \pp and \PbPb collisions, respectively, and the \raa of prompt \PDz{} mesons are presented as a function of their \pt. The \raa is reported for two centrality intervals: in the inclusive sample (0--100\%), and in one corresponding to the most overlapping 10\% of the collisions.

\section{The CMS detector}
The central feature of the CMS apparatus is a superconducting solenoid of 6\unit{m} internal diameter, providing a magnetic field of 3.8\unit{T}.
Within the solenoid volume are a silicon tracker which measures charged particles within the pseudorapidity range $\abs{\eta} < 2.5$, a lead tungstate crystal electromagnetic calorimeter (ECAL), and a brass and scintillator hadron calorimeter (HCAL). The ECAL consists of more than 75\,000 lead tungstate crystals, and is partitioned into a barrel region ($\abseta < 1.48$) and two endcaps extending out to $\abseta = 3.0$. The HCAL consists of sampling calorimeters composed of brass and scintillator plates, covering $\abseta < 3.0$. Iron hadron forward (HF) calorimeters, with quartz fibers read out by photomultipliers, extend the calorimeter coverage out to $\abseta = 5.2$. A detailed description of the CMS experiment can be found in Ref.~\cite{bib_CMS}.

\section{Event selection and Monte Carlo samples}
The \pp (\PbPb{}) dataset used for this analysis corresponds to an integrated luminosity of 27.4\pbinv (530\mubinv). The \PDz{} meson production is measured from \pt of 2 up to 20\GeVc using large samples of minimum-bias (MB) events ($\approx$2.5 billion \pp events and $\approx$300 million \PbPb events).
Minimum-bias events were selected online using the information from the HF calorimeters and the beam pickup monitors. For measuring the \PDz{} meson production above 20\GeVc, dedicated high-level trigger (HLT) algorithms were designed to identify online events with a \PDz{} candidate. Since events with a high-\pt \PDz{} meson are expected to leave large energy deposits in HCAL, HLT algorithms were run on events preselected by jet triggers in the level-1 (L1) calorimeter trigger system. In \PbPb collisions, the \PDz{} triggers with \pt threshold below 40\GeVc were run on events passing the L1 MB trigger selection. While the MB and lower-threshold triggers had to be prescaled because of the high instantaneous luminosity of the LHC, the highest threshold trigger used in the analysis ($\pt>60\,(50)$\GeVc for \PbPb (\pp) data taken) was always unprescaled. The efficiency of the HLT algorithms was evaluated in data, and modeled by a linear function of \PDz{} \pt. The efficiency was found to be about 100 (90)\% in \pp (\PbPb{}) collisions for events passing the corresponding L1 selection.

For the offline analysis, events have to pass a set of selection criteria designed to reject events from background processes (beam-gas collisions and beam scraping events) as described in Ref.~\cite{Khachatryan:2016odn}. In order to select hadronic collisions, both \pp and \PbPb events are required to have at least one reconstructed primary interaction vertex with a distance from the centre of the nominal interaction region of less than 15\cm along the beam axis. In addition, in \PbPb collisions the shapes of the clusters in the pixel detector have to be compatible with those expected from particles produced by a \PbPb
collision~\cite{Khachatryan:2010xs}. The \PbPb collision events are also required to have at least three towers in each of the HF detectors with energy deposits of more than 3\GeV per tower. The combined efficiency for this event selection, and the remaining non-hadronic contamination, is $(99\pm2)$\%. Selection efficiencies higher than 100\% are possible, reflecting the possible presence of ultra-peripheral (nonhadronic) collisions in the selected event sample.
The collision centrality is determined from the total transverse energy deposition in both the HF calorimeters. Collision centrality bins are given in percentage ranges of the total inelastic hadronic cross section, with the 0--10\% bin corresponding to the 10\% of collisions having the largest overlap of the two nuclei.

Several Monte Carlo (MC) simulated event samples are used to evaluate background components, signal efficiencies, and detector acceptance corrections. The events produced include both prompt and nonprompt (from b hadron decays) \PDz{} meson events. Proton-proton collisions are generated with \PYTHIA 8 v212~\cite{Sjostrand:2014zea} tune CUETP8M1~\cite{Khachatryan:2015pea} and propagated through the CMS detector using the \GEANTfour package~\cite{geant4}. The \PDz{} mesons are decayed with \EVTGEN 1.3.0~\cite{evtgen}, and final-state photon radiation in the \PDz{} decays is simulated with \PHOTOS 2.0~\cite{Barberio:1990ms}. For the $\PbPb$ MC samples, each \PYTHIA 8 event is embedded into a $\PbPb$ collision event generated with \textsc{hydjet}~1.8~\cite{Lokhtin:2005px}, which is tuned to reproduce global event properties such as the charged-hadron \pt spectrum and particle multiplicity.

\section{Signal extraction}
\label{sec:SignalExtraction}

The \PDz{} candidates are reconstructed by combining pairs of oppositely charged particle tracks with an invariant mass within
0.2\GeVcc of the world-average \PDz{} mass~\cite{pdg:2016}. Each track is required to have $\pt>1$\GeVc in order to reduce the combinatorial background. For high-\pt \PDz{} mesons (above 20\GeVc) in \PbPb data, the single track cut is raised to $\pt >8.5$\GeVc to account for the selection ($\pt>8$\GeVc) performed at the HLT. All tracks are also required to be within $\abseta < 1.5$. For each pair of selected tracks, two \PDz{} candidates are created by assuming that one of the particles has the mass of the pion while the other has the mass of the kaon, and vice-versa.
The \PDz{} mesons are required to be within $\absy< 1$, optimised in conjunction to the track pseudorapidity selection to give the best signal to background ratio over the whole range of \PDz{} \pt studied. In order to further reduce the combinatorial background, the \PDz{} candidates are selected based on three topological criteria: on the three-dimensional (3D) decay length ${L_\mathrm{xyz}}$ normalised to its uncertainty (required to be larger than 4--6), on the pointing angle $\theta_\mathrm{p}$ (defined as the angle between the total momentum vector of the tracks and the vector connecting the primary and the secondary vertices and required to be smaller than 0.12), and on the $\chi^{2}$ probability, divided by the number of degrees of freedom, of the \PDz{} vertex fit (required to be larger than 0.025--0.05).
The selection is optimised in each \pt bin using a multivariate technique~\cite{Hocker:2007ht} in order to maximise the statistical significance of the \PDz{} meson signals.

\begin{figure*}[htb]
\centering
\includegraphics[width= 0.44\textwidth]{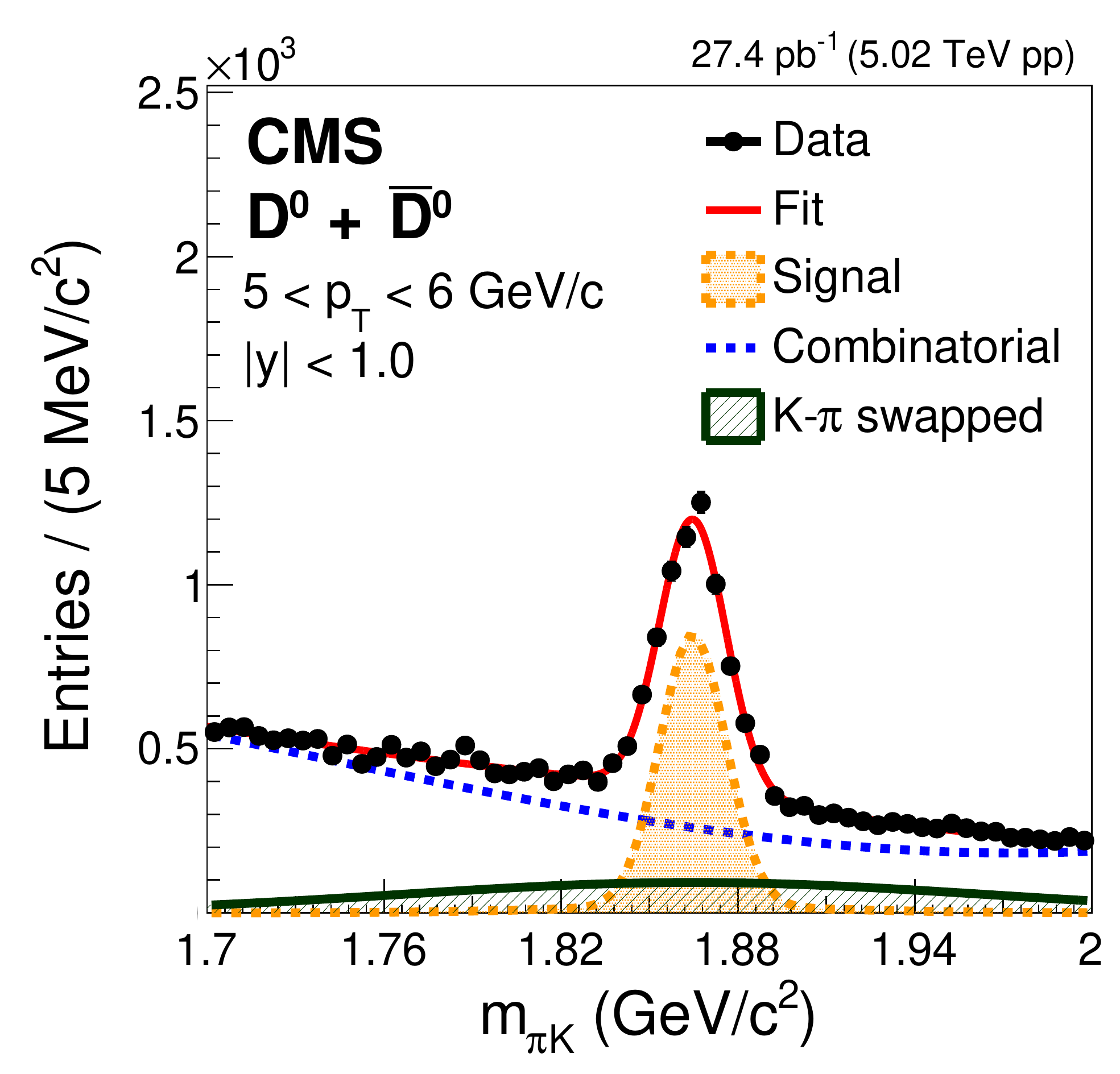}
\includegraphics[width= 0.44\textwidth]{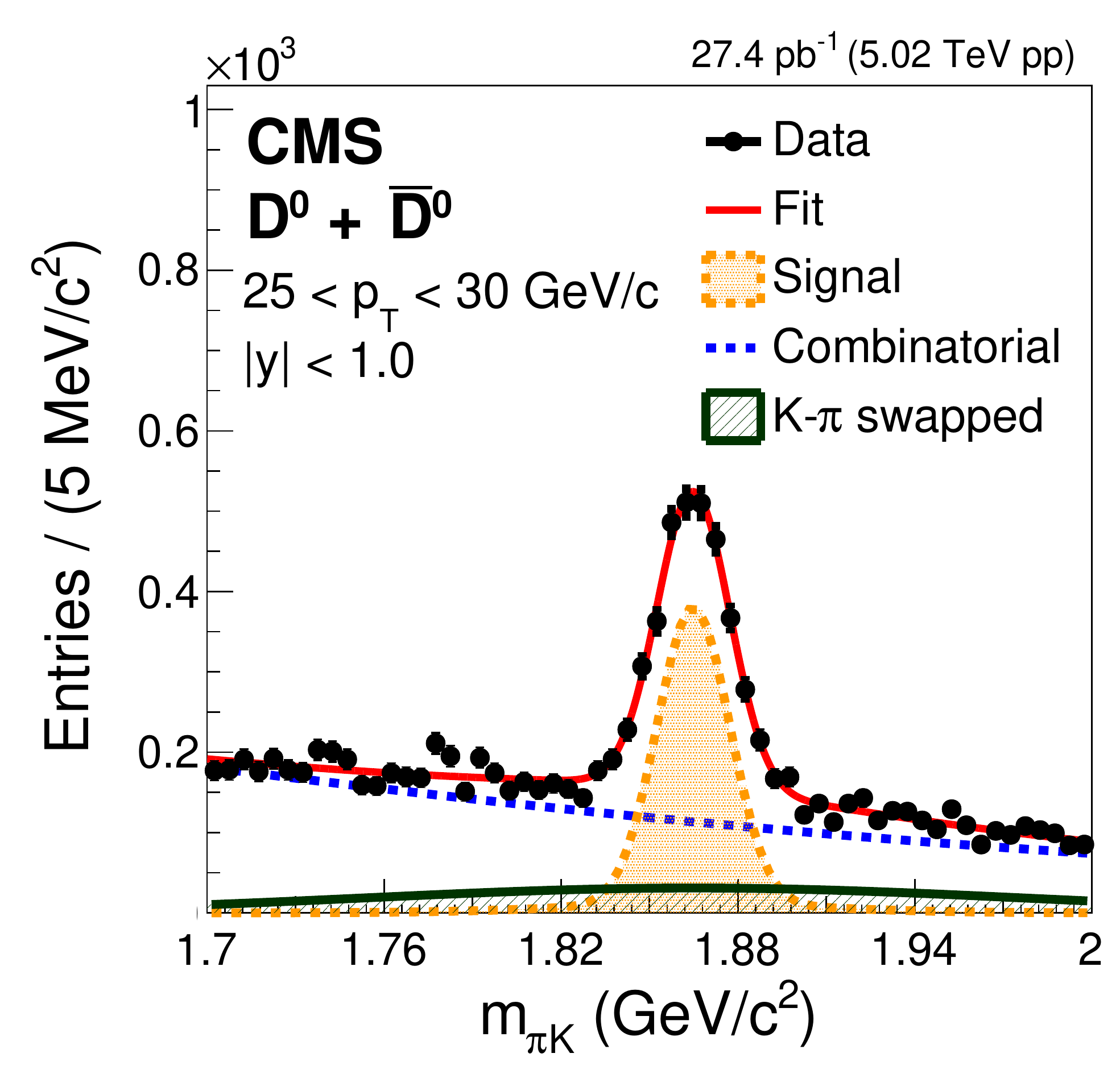}
\includegraphics[width= 0.44\textwidth]{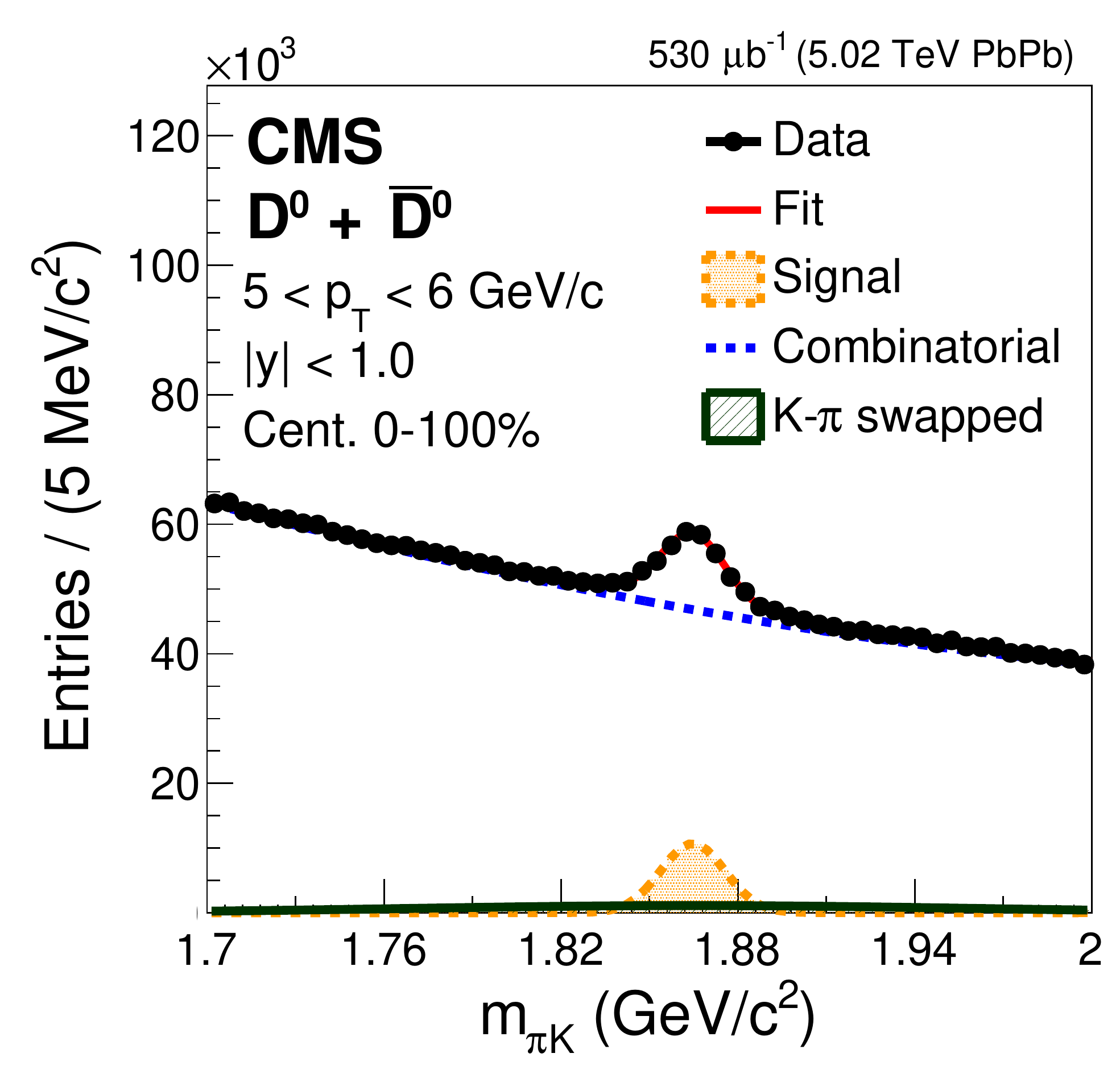}
\includegraphics[width= 0.44\textwidth]{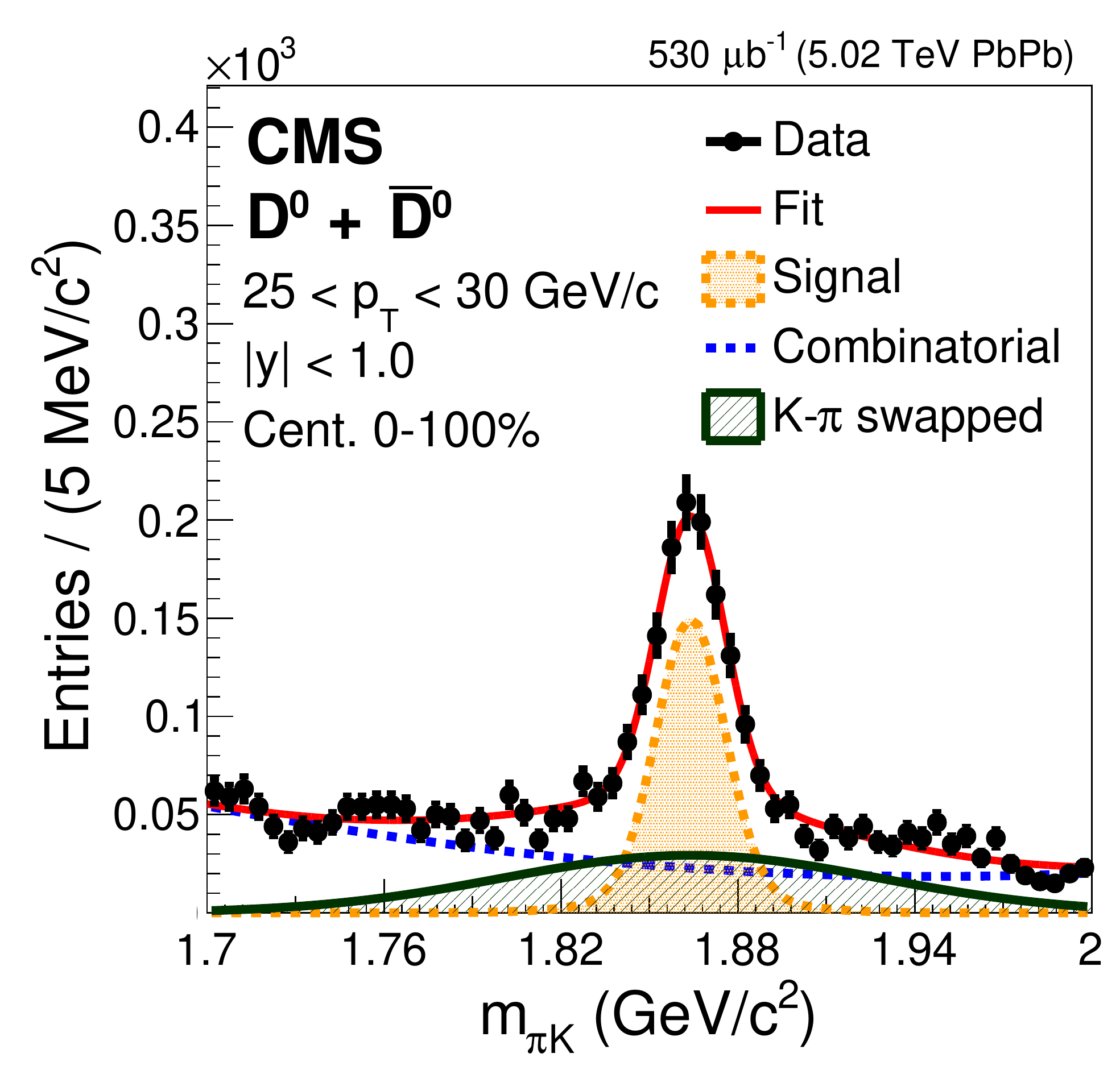}
\caption{Examples of \PDz{} candidate invariant mass distributions in \pp (top) and \PbPb (bottom) collisions at 5.02\TeV.}
\label{fig:yieldsDzero}
\end{figure*}

The \PDz{} meson yields in each $\pt$ interval are extracted with a binned maximum-likelihood fit to the invariant mass distributions in the range $1.7 <m_{\PGp \PK}< 2.0$\GeVcc. Several examples of \PDz{} candidate invariant mass distributions are shown in Fig.~\ref{fig:yieldsDzero} for \pp (top) and \PbPb (bottom) collisions. The combinatorial background, originating from random pairs of tracks not produced by a \PDz{} meson decay, is modeled by a third-order polynomial. The signal shape was found to be best modeled over the entire \pt range measured by two Gaussian functions with the same mean but different widths. An additional Gaussian function is used to describe the invariant mass shape of \PDz{} candidates with incorrect mass assignment from the exchange of the pion and kaon designations. The widths of the Gaussian functions that describe the \PDz{} signal shape and the shape of the \PDz{} candidates with swapped mass assignment are free parameters in the fit. Also, the ratio between the yields of the signal and of the  \PDz{} candidates with swapped mass assignments is fixed to the value extracted from simulation.

The \PDz{} \pt-differential cross section in each \pt interval in \pp collisions is defined as:
\begin{equation}
  \left.\frac{\rd{}\sigma_{\pp}}{\rd{}\pt}\right|_{\absy < 1} =
  \frac{1}{2}\frac{1}{\Delta\pt}\frac{1}{\mathcal{B} \, \mathcal{L}} \left.\frac{f_\text{prompt} \, N_{\pp} }{( \alpha \,\epsilon)_\text{prompt} \, \beta_\text{prescale} \,\epsilon_\text{trigger}}\right|_{\absy<1},
  \label{eq:crosssectionPP}
\end{equation}
where $\Delta \pt$ is the width of the $\pt$ interval, $\mathcal{B}$ is the branching fraction of the decay chain, $\mathcal{L}$ is the integrated luminosity,  $(\alpha \,\epsilon)_\text{prompt}$ represents the correction for acceptance and efficiency and $N_{\pp}$ is the yield of $\PDz$ and $\PADz$ mesons extracted in each \pt interval. In both \pp and \PbPb cases, the value of $\alpha_\text{prompt}$ ranges from about 0.3 at 2--3\GeVc to about 100\% at 60--100\GeVc. The value of $\epsilon_\text{prompt}$ ranges for \PbPb (\pp) from about 0.02 (0.03) at 2--3\GeVc to about 0.4 (0.6) at 60--100\GeVc. The factor 1/2 accounts for the fact that the cross section is given for the average of particles and antiparticles. The raw yields $N_{\pp}$ are corrected in order to account for the average prescale factor $\beta_\text{prescale}$ and the efficiency $\epsilon_\text{trigger}$ of the trigger that was used to select events in that specific \pt interval.
The factor $f_\text{prompt}$ is the fraction of \PDz{} mesons that comes directly from c quark fragmentation and is measured using control samples in data by exploiting the difference in the distributions of a quantity found by multiplying the 3D \PDz{} decay length $L_{xyz}$ by the sine of the pointing angle $\sin(\theta_\mathrm{p})$ of prompt and nonprompt \PDz{} mesons. In particular, the value of $f_\text{prompt}$ (typically in the range 0.8--0.9) is measured in each \pt interval by fitting the distribution of ${L_\mathrm{xyz}} \sin(\theta_\mathrm{p})$ using the prompt and nonprompt shapes obtained from MC simulation.

The \PDz{} \pt-differential production yield in each \pt interval in \PbPb collisions is defined as:
\ifthenelse{\boolean{cms@external}}{
\begin{multline}
 \frac{1}{\taa} \left.\frac{\rd{}N_{\PbPb}}{\rd{}\pt}\right|_{\absy < 1} \\ =
 \frac{1}{\taa}\frac{1}{2}\frac{1}{\Delta\pt}\frac{1}{\mathcal{B} \, N_{\text{MB}}} \left.\frac{f_\text{prompt} \, N_{\PbPb}}{( \alpha \,\epsilon)_\text{prompt} \, \beta_\text{prescale} \,~\epsilon_\text{trigger}}\right|_{\absy<1},
\label{eq:PbPbYield}
\end{multline}}{
\begin{equation}
 \frac{1}{\taa} \left.\frac{\rd{}N_{\PbPb}}{\rd{}\pt}\right|_{\absy < 1} =
 \frac{1}{\taa}\frac{1}{2}\frac{1}{\Delta\pt}\frac{1}{\mathcal{B} \, N_{\text{MB}}} \left.\frac{f_\text{prompt} \, N_{\PbPb}}{( \alpha \,\epsilon)_\text{prompt} \, \beta_\text{prescale} \,~\epsilon_\text{trigger}}\right|_{\absy<1},
\label{eq:PbPbYield}
\end{equation}
}
where $N_\mathrm{MB}$ is the number of MB events used for the analysis and \taa is the nuclear overlap function~\cite{Miller:2007ri}, which is equal to the number of nucleon-nucleon (NN) binary collisions divided by the NN cross section and can be interpreted as the NN-equivalent integrated luminosity per heavy ion collision. The values of \taa are 5.61\mbinv for inclusive \PbPb collisions and 23.2\mbinv  for central events~\cite{Khachatryan:2016odn}.
The other terms were defined analogously to Eq.~(\ref{eq:crosssectionPP}).

 \section{Systematic uncertainties}
The yields are affected by several sources of systematic uncertainties arising from the signal extraction, acceptance and efficiency corrections, branching fraction, and integrated luminosity determination.
The uncertainty in the raw yield extraction (1.6--8.2\% for \pp and 1.3--17.5\% for \PbPb data, with the highest value at low-\pt, which is the region with the smallest signal to background ratio) is evaluated by repeating the
fit procedure using different background fit functions and by forcing the widths of the Gaussian functions that describe the signal to be equal to the values extracted in simulations to account for possible differences in the signal resolution in data and in MC.
In the background variation study, an exponential plus a second-order polynomial function was considered instead of the first order polynomial one, which is used as default.
The final uncertainty in the raw yield extraction is defined as the sum in quadrature of the relative differences of the signal variation and the maximum of all the background variations.

The systematic uncertainty due to the selection of the \PDz{} meson candidates (0.5--3.6\% for \pp and 2.7--8.1\% for \PbPb data, with the highest value at low-\pt) is estimated by considering the differences between MC and data in the reduction
of the \PDz{} yields obtained by applying each of the \PDz{} selection variables described in Sec.~\ref{sec:SignalExtraction}.
The study was performed by varying one selection at a time, in a range that allowed a robust signal extraction procedure and by considering the maximum relative discrepancy in the yield reduction between data and MC.
The total uncertainty was the quadratic sum of the maximum relative discrepancy obtained by varying each of the three selection variables separately.

The uncertainty due to the \PDz{} trigger efficiency (1\% for \pp and 2\% for \PbPb data) is evaluated as the statistical uncertainty in the zeroth-order coefficient of the linear function used to describe the plateau of the efficiency distribution.
The systematic uncertainty in the hadron tracking efficiency (4.0\% for \pp and 6.0--6.5\% for \PbPb data) is estimated from a comparison of two- and four-body \PDz{} meson decays in data and simulated samples~\cite{CMS-PAS-TRK-10-002}.

To evaluate the systematic uncertainty in the prompt \PDz{} meson fraction, the width of the ${L_\mathrm{xyz}} \sin(\theta_\mathrm{p})$ MC prompt and nonprompt templates are varied in a range that covers the observed differences between
the data and MC values. The systematic uncertainty (10\% for both \pp and \PbPb data) was obtained in each \pt bin as the difference between the $f_\text{prompt}$ value extracted from the variation that gives the best $\chi^2$ fit to data and the nominal $f_\text{prompt}$ value.
To evaluate this uncertainty for the \raa measurement, the widths of the template distributions are varied simultaneously in \pp and \PbPb. The systematic uncertainty on the $f_\text{prompt}$ correction was evaluated as the spread of the ratios
of $f_\text{prompt}$ in \PbPb and \pp to account for partial cancellations of the systematic effects in the two analyses.

The uncertainty related to the simulated \pt shape (smaller than 0.5\% for both \pp and \PbPb data) is evaluated by reweighting the simulated \PDz{} meson \pt distribution according to the \pt shape obtained from a fixed-order plus next-to-leading logarithmic (FONLL) prediction~\cite{Cacciari:1998it}.

The systematic uncertainty in the cross section measurement is computed as the sum in quadrature of the different contributions mentioned above.
The global uncertainty in the \pp measurement (2.5\%) is the sum in quadrature of the systematic uncertainty in the integrated luminosity (2.3\%~\cite{CMS-PAS-LUM-16-001}) and in the branching fraction $\mathcal{B}$ (1.0\%~\cite{pdg:2016}).
The global uncertainty in the \PbPb measurement ($+$3.6\%, $-$4.1\% for the centrality range 0--100\% and $+$2.9\%, $-$3.7\% for 0--10\%) is the sum in quadrature of the uncertainties in the MB selection efficiency (2\%),
in the branching fraction (1.0\%) and in the \taa ($+$2.8\%, $-$3.4\% for the centrality range 0--100\% and $+$1.9\%, $-$3.0\% for 0--10\%). For the \raa results, no cancelation of uncertainties is assumed between the \pp and \PbPb results.

\section{Results}
\begin{figure*}[htb]
\centering
\includegraphics[width= 0.49\textwidth]{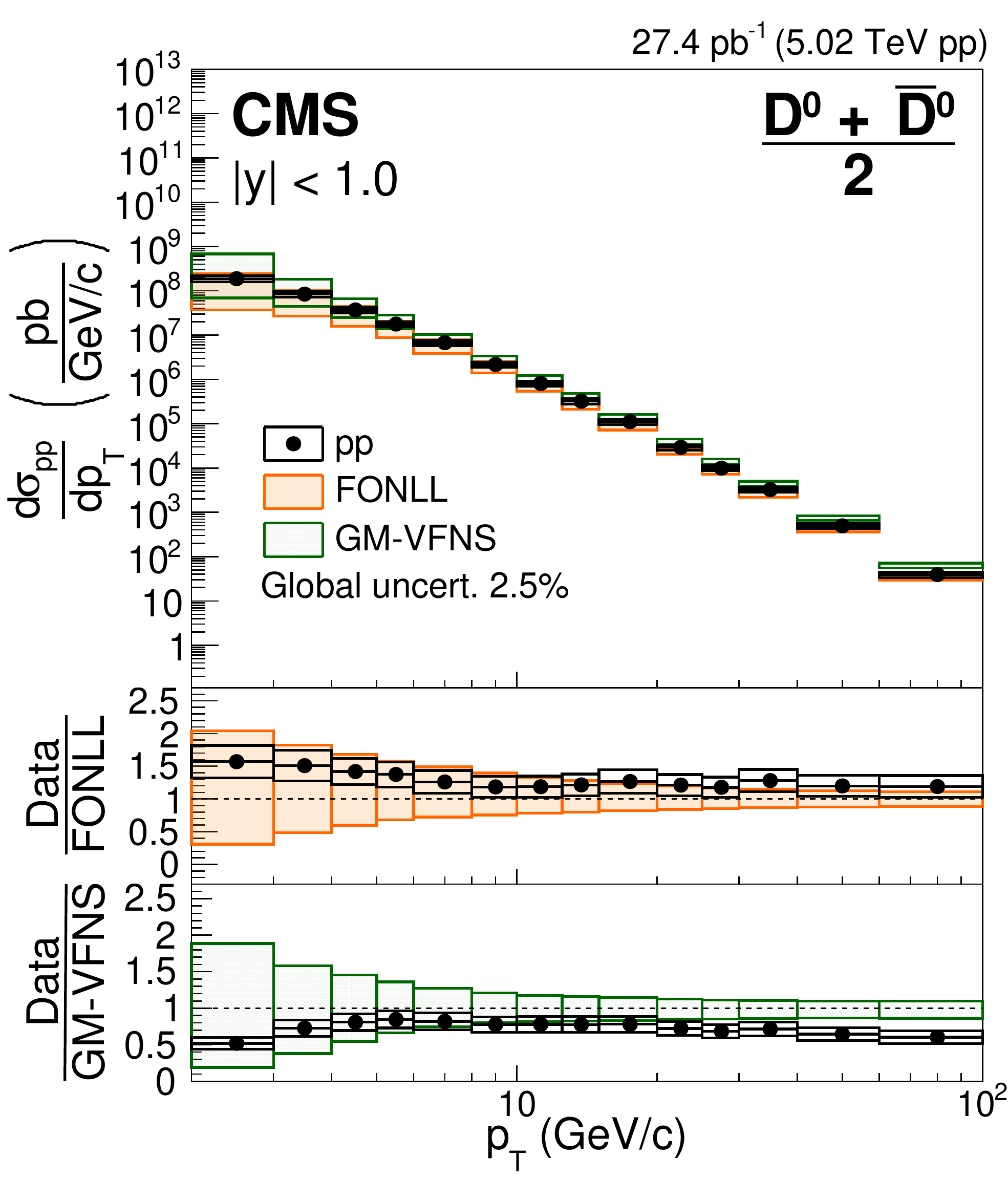}
\includegraphics[width= 0.49\textwidth]{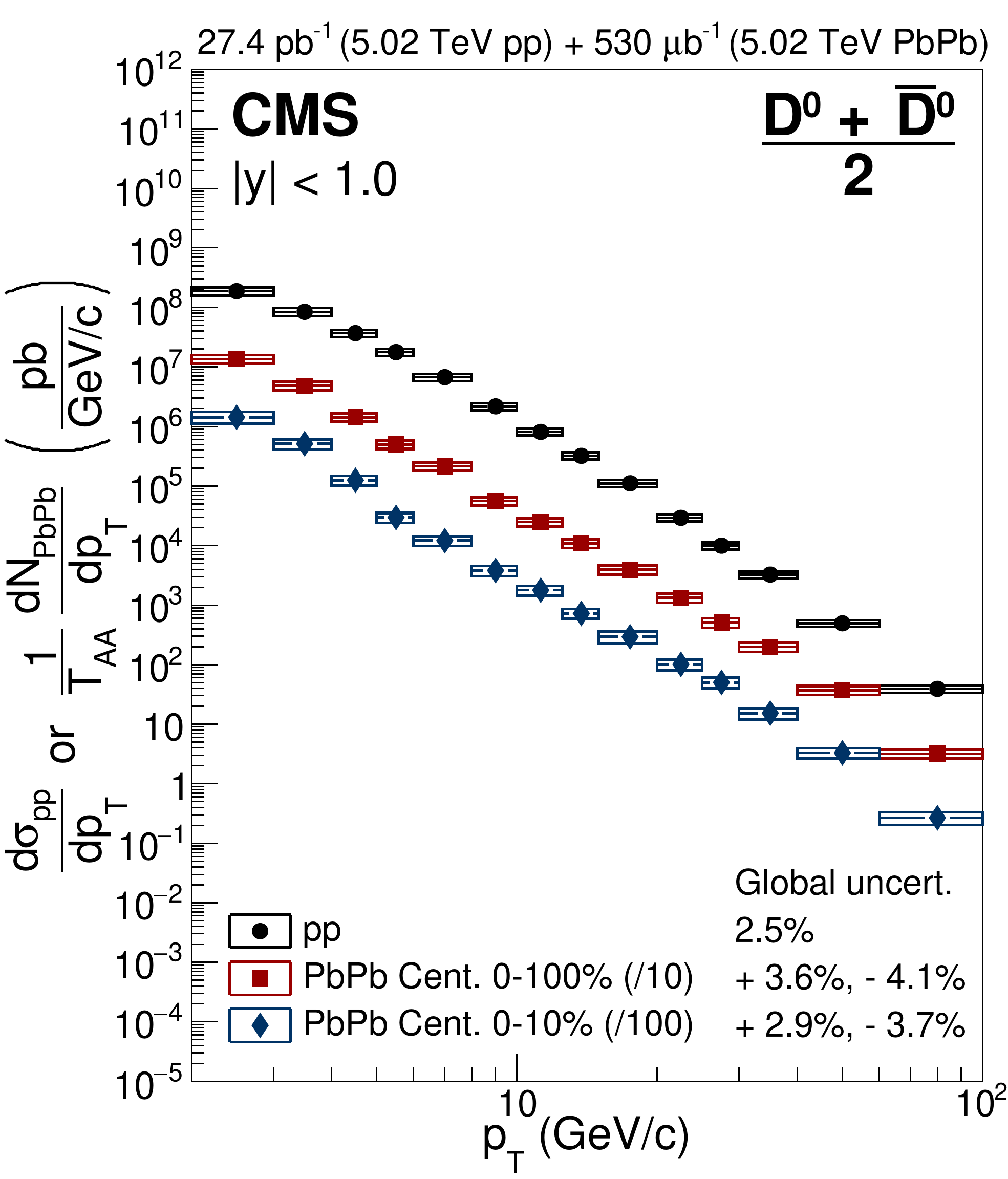}
\caption{(left) The $\pt$-differential production cross section of $\PDz$ mesons in \pp collisions at $\sqrts=5.02\TeV$. The vertical bars (boxes) correspond to statistical (systematic) uncertainties.
The global systematic uncertainty, listed in the legend and not included in the point-to-point uncertainties, comprises the uncertainties in the integrated luminosity measurement and the \PDz{} meson $\mathcal{B}$.
Results are compared to FONLL ~\cite{Cacciari:1998it} and GM-VFNS~\cite{Kniehl:2004fy,Kniehl:2005mk,Kniehl:2012ti} calculations. (right) The \pt-differential production yields of \PDz{} mesons divided by the nuclear overlap functions \taa
for \PbPb collisions in the 0--100\% (red) and  0--10\% (blue) centrality ranges compared to the same \pp cross sections shown in the left panel (black).}
\label{fig:crosssectionPPcombined}
\end{figure*}

The $\pt$-differential production cross section in \pp collisions measured in the interval $\absy< 1$ is presented in the left panel of Fig.~\ref{fig:crosssectionPPcombined}.
The result is compared to the prediction of FONLL and a general-mass variable flavour number scheme (GM-VFNS)~\cite{Kniehl:2004fy,Kniehl:2005mk,Kniehl:2012ti} calculation. The CMS measurement lies close to the upper bound of the FONLL prediction and the lower bound of the GM-VFNS calculation. The \PDz{} \pt-differential production yields divided by the nuclear overlap functions \taa in \PbPb collisions in the 0--100\% and  0--10\% centrality ranges are presented in the right panel of Fig.~\ref{fig:crosssectionPPcombined} and compared to the same \pp cross section shown in the left panel.

\begin{figure}[htbp]
\centering
\includegraphics[width= 0.49\textwidth]{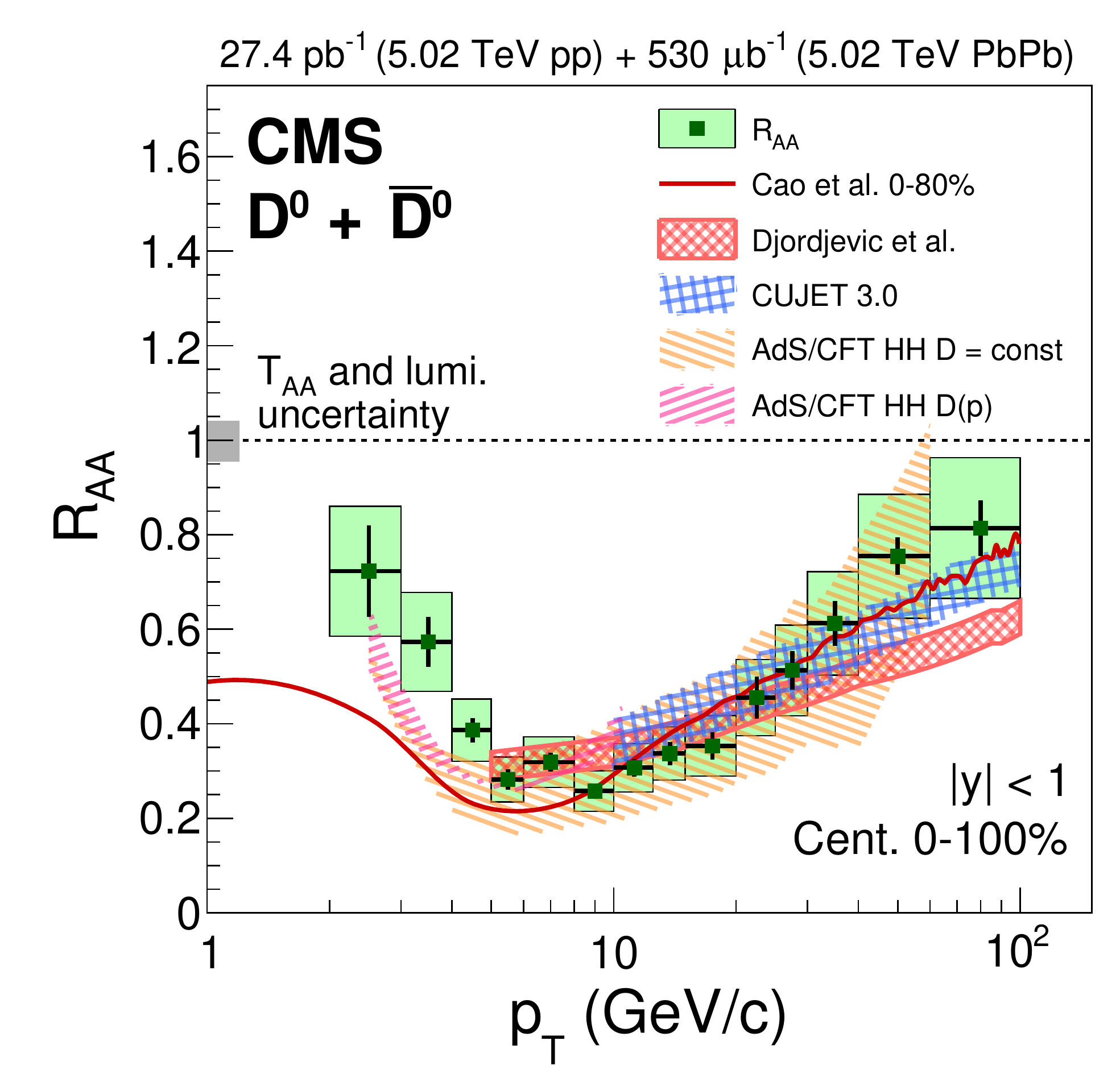}
\includegraphics[width= 0.49\textwidth]{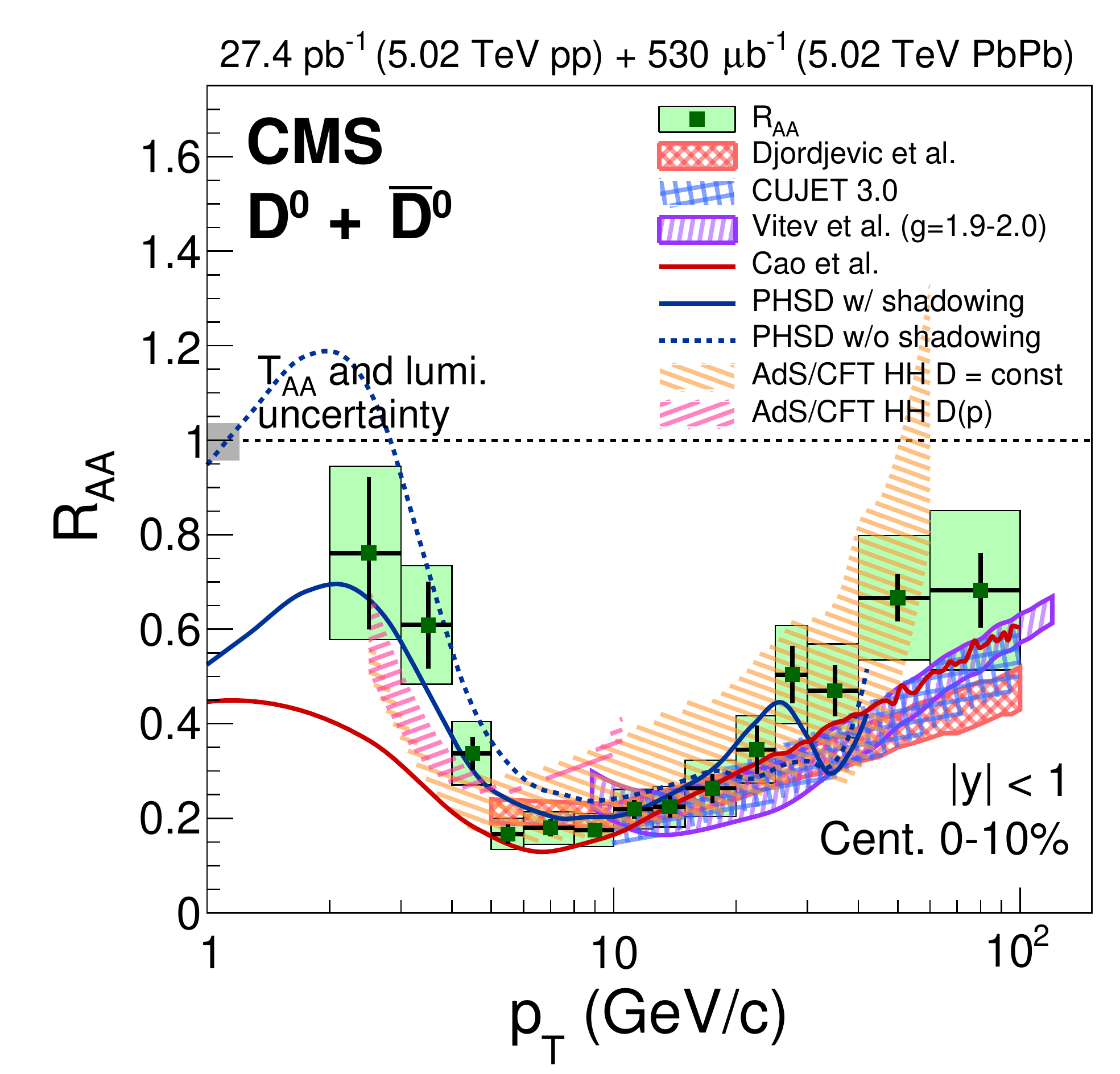}
\caption{\raa as a function of \pt in the centrality range 0--100\% (\cmsLeft) and  0--10\% (\cmsRight).
The vertical bars (boxes) correspond to statistical (systematic) uncertainties.
The global systematic uncertainty, represented as a grey box at $\raa=1$, comprises the uncertainties in the integrated luminosity measurement and \taa value.
The \PDz{} \raa values are also compared to calculations from various theoretical models~\cite{Djordjevic:2015hra,Xu:2015bbz,Xu:2014ica,Xu:2014tda,Kang:2014xsa,Chien:2015vja,Cao:2016gvr,Cao:2017hhk,Song:2015sfa, Song:2015ykw,Horowitz:2015dta}.}
\label{fig:NuclearModificationTheory}
\end{figure}

The nuclear modification factor, \raa is computed as:
\begin{equation}
 \raa=\left.\frac{1}{\taa} \frac{\rd{}N_{\PbPb}}{\rd{}\pt} \middle/ \frac{\rd{}\sigma_{\pp}}{\rd{}\pt}\right..
  \label{eq:RAA}
\end{equation}
The \raa in the centrality range 0--100\% is shown in the left panel of Fig.~\ref{fig:NuclearModificationTheory} as a function of \pt. The \raa shows a suppression of a factor 3 to 4 at \pt of 6--8\GeVc. At higher \pt, the suppression factor decreases to a value of about 1.3 in the \pt range 60--100\GeVc. The \raa for the centrality range 0--10\% is presented in the right panel of Fig.~\ref{fig:NuclearModificationTheory}.
The \PDz{} \raa in central events shows a hint of stronger suppression if compared to the inclusive \raa result for $\pt>5$\GeVc.
In this comparison, the large overlap between the two results has to be considered. Indeed, roughly 40$\%$ of the \PDz{} candidates used in the measurement in the centrality range 0--100$\%$ are
also included in the 0--10$\%$ result.

The results are also compared to calculations of four types of models: (a) two perturbative QCD-based models that include both collisional and radiative energy loss, (M. Djordjevic \cite{Djordjevic:2015hra} and CUJET 3.0~\cite{Xu:2015bbz, Xu:2014ica,Xu:2014tda}) and one that includes radiative energy loss only (I. Vitev \cite{Kang:2014xsa,Chien:2015vja}), (b) a transport model based on a Langevin equation that includes collisional energy loss and heavy-quark diffusion in the medium (S. Cao et~al. \cite{Cao:2016gvr,Cao:2017hhk}), (c) a microscopic off-shell transport model based on a Boltzmann approach that includes collisional energy loss only (PHSD~\cite{Song:2015sfa, Song:2015ykw}), and (d) a model based on the anti-de Sitter/conformal field theory (AdS/CFT) correspondence, that includes thermal fluctuations in the energy loss for heavy quarks in a strongly coupled plasma~\cite{Horowitz:2015dta}. The AdS/CFT calculation is provided for two settings of the diffusion coefficient D of the heavy quark propagation through the medium: dependent on, and independent of the quark momentum.
For \PDz{} meson $\pt >40$\GeVc, the perturbative QCD-based models describe the suppression in both centrality ranges within the uncertainties, although the trend suggested by these predictions is typically lower than that in the experimental data.
The model based on a Langevin approach describes the measurement well in the centrality range 0--100\%, while it predicts slightly too much suppression for central events. The AdS/CFT calculations describe well both the 0--100\% and the 0--10\% measurements. In the intermediate \pt region ($10 < \pt < 40$\GeVc), all the theoretical calculations describe well the \raa results in both centrality intervals. For $\pt < 10$\GeVc, the PHSD prediction that includes shadowing can reproduce the measurement in the 0--100\% centrality region accurately, while the Langevin calculation predicts significantly more suppression than seen in data for both centrality ranges. In the same low-\pt region, the AdS/CFT calculation lies at the lower limit of the experimental uncertainties for both 0--10\% and 0--100\% measurements.

\begin{figure}[htbp]
\centering
\includegraphics[width= 0.49\textwidth]{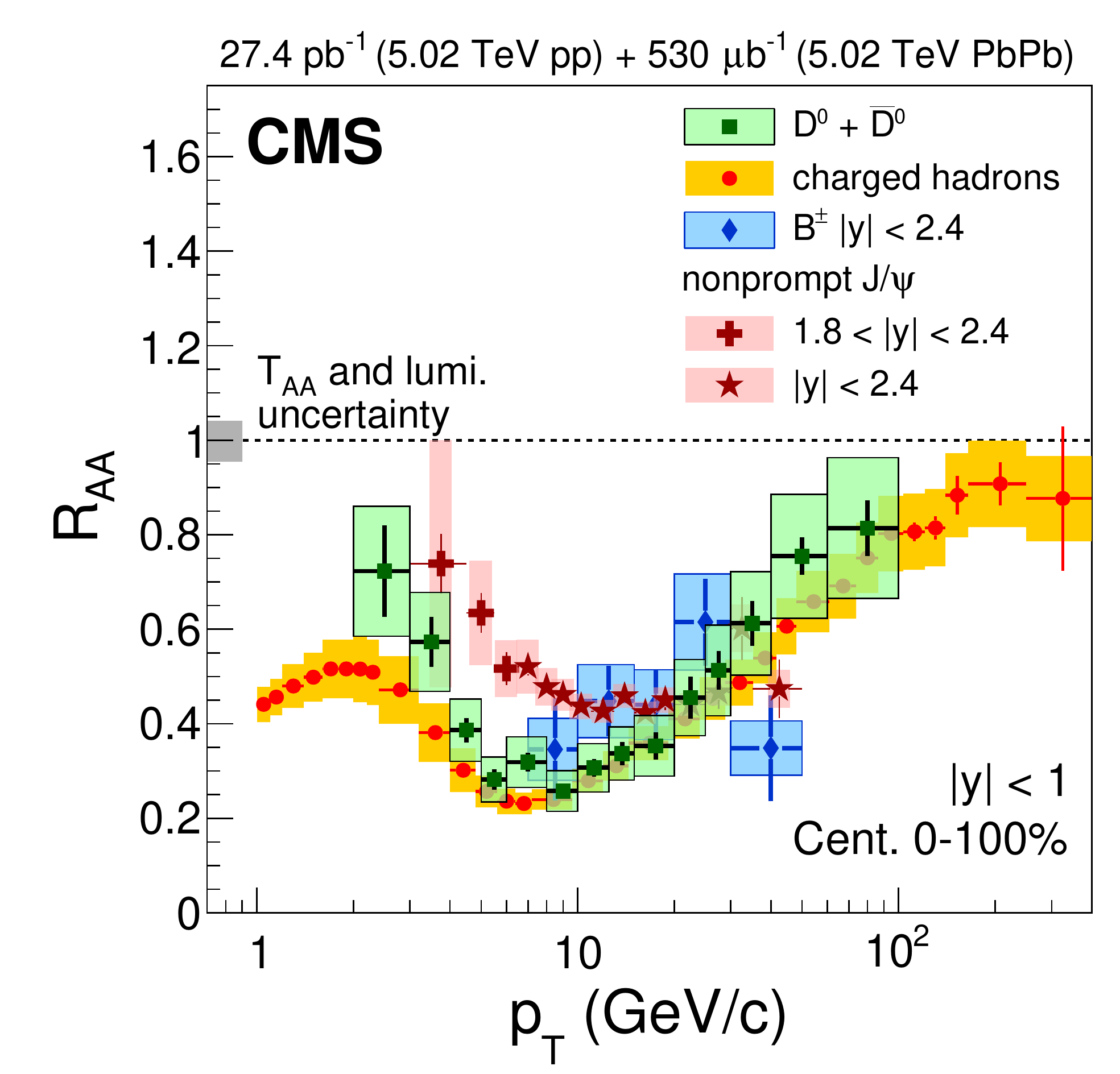}
\includegraphics[width= 0.49\textwidth]{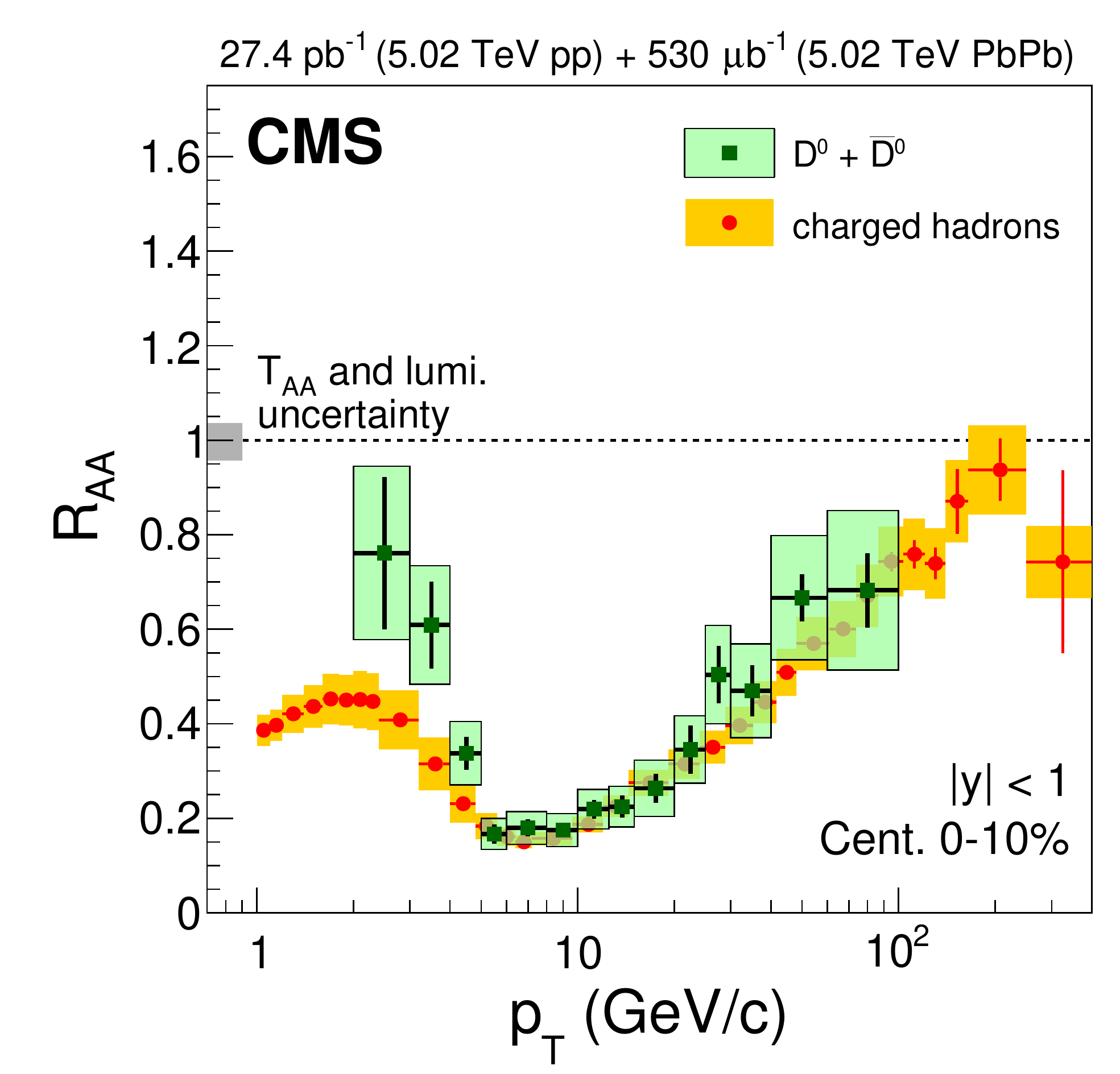}
\caption{(\cmsLeft) Nuclear modification factor \raa as a function of \pt in the centrality range 0--100\% (green squares) compared to the \raa of charged particles (red circles)~\cite{Khachatryan:2016odn},
\Bplusminus mesons (blue triangles)~\cite{Bmesonpaper} and nonprompt \PJGy meson (purple crosses and stars)~\cite{Nonprompt5TeV} in the same centrality range at 5.02\TeV.
(\cmsRight) Nuclear modification factor \raa as a function of \pt in the centrality range 0--10\% (green squares) compared to the \raa of charged particles (red circles) ~\cite{Khachatryan:2016odn} in the
same centrality range. }
\label{fig:NuclearModification}
\end{figure}

The \PDz{} \raa measured in the centrality range 0--100\% is compared in the top panel of Fig.~\ref{fig:NuclearModification} to the CMS measurements of the \raa of charged particles~\cite{Khachatryan:2016odn}, \Bplusminus mesons~\cite{Bmesonpaper} and nonprompt \PJGy meson~\cite{Nonprompt5TeV} performed at the same energy and in the same centrality range. The systematic uncertainties between the \raa measurement of the \PDz{} mesons, and of the light and beauty particles, are almost completely uncorrelated. The only common contribution comes from the systematic uncertainty of one track (4\%), which is however negligible when compared to the total uncertainties. The \PDz{} meson \raa values are consistent with those of charged particles for $\pt > 4$\GeVc. For lower \pt, a somewhat smaller suppression for \PDz{} mesons is observed. The \raa of the \Bplusminus mesons, measured in the \pt range 7--50\GeVc and the rapidity range of $\absy < 2.4$, is also consistent with the \PDz{} meson measurement within the experimental uncertainties. The \raa of nonprompt \PJGy, which was found to have almost no rapidity dependence~\cite{Nonprompt5TeV}, is shown here measured in the \pt ranges 6.5--50\GeVc in $\absy < 2.4$, and 3--6.5\GeVc in $1.8< \absy < 2.4$. Its \raa is found to be higher than the \PDz{} meson \raa in almost the entire \pt range. The \PDz{} meson \raa in the centrality range 0--10\% is compared in Fig.~\ref{fig:NuclearModification} to the charged-particle \raa. As observed for 0--100\% \PbPb events, the two results
are consistent within uncertainties for $\pt > 4$\GeVc and a somewhat smaller suppression for charmed mesons is observed at lower \pt.

\section{Summary}
In this Letter, the transverse momentum (\pt) spectra of prompt \PDz{} mesons in \pp and \PbPb collisions and the \PDz{} meson nuclear modification factor (\raa) in the central rapidity region ($\absy< 1$) at $\sqrtsNN= 5.02$\TeV from CMS are presented. The \raa of prompt \PDz{} mesons is measured as a function of their \pt from 2 to 100\GeVc in two centrality ranges, inclusive and 10\% most central. The \PDz{} meson yield is found to be strongly suppressed in \PbPb collisions when compared to the measured \pp reference data scaled by the number of binary nucleon-nucleon collisions. These measurements are consistent with the \raa of charged hadrons in both centrality intervals for $\pt > 4$\GeVc. A hint of a smaller suppression of \PDz{} \raa with respect to charged particle \raa is observed for $\pt < 4$\GeVc. The \PDz{} \raa was found to be compatible with the \Bplusminus \raa in the intermediate \pt region and significantly lower than the nonprompt \PJGy meson \raa for $\pt< 10$\GeVc. Comparisons to different theoretical models show that the general trend of the \raa is qualitatively reproduced at high \pt.
Comparisons to different theoretical models show that the general trend of the \raa is qualitatively reproduced at high \pt, while quantitative agreement for all centrality and \pt selections is yet to be attained.

\begin{acknowledgments}
We congratulate our colleagues in the CERN accelerator departments for the excellent performance of the LHC and thank the technical and administrative staffs at CERN and at other CMS institutes for their contributions to the success of the CMS effort. In addition, we gratefully acknowledge the computing centres and personnel of the Worldwide LHC Computing Grid for delivering so effectively the computing infrastructure essential to our analyses. Finally, we acknowledge the enduring support for the construction and operation of the LHC and the CMS detector provided by the following funding agencies: BMWFW and FWF (Austria); FNRS and FWO (Belgium); CNPq, CAPES, FAPERJ, and FAPESP (Brazil); MES (Bulgaria); CERN; CAS, MoST, and NSFC (China); COLCIENCIAS (Colombia); MSES and CSF (Croatia); RPF (Cyprus); SENESCYT (Ecuador); MoER, ERC IUT, and ERDF (Estonia); Academy of Finland, MEC, and HIP (Finland); CEA and CNRS/IN2P3 (France); BMBF, DFG, and HGF (Germany); GSRT (Greece); OTKA and NIH (Hungary); DAE and DST (India); IPM (Iran); SFI (Ireland); INFN (Italy); MSIP and NRF (Republic of Korea); LAS (Lithuania); MOE and UM (Malaysia); BUAP, CINVESTAV, CONACYT, LNS, SEP, and UASLP-FAI (Mexico); MBIE (New Zealand); PAEC (Pakistan); MSHE and NSC (Poland); FCT (Portugal); JINR (Dubna); MON, RosAtom, RAS, RFBR and RAEP (Russia); MESTD (Serbia); SEIDI, CPAN, PCTI and FEDER (Spain); Swiss Funding Agencies (Switzerland); MST (Taipei); ThEPCenter, IPST, STAR, and NSTDA (Thailand); TUBITAK and TAEK (Turkey); NASU and SFFR (Ukraine); STFC (United Kingdom); DOE and NSF (USA).

\hyphenation{Rachada-pisek} Individuals have received support from the Marie-Curie programme and the European Research Council and Horizon 2020 Grant, contract No. 675440 (European Union); the Leventis Foundation; the A. P. Sloan Foundation; the Alexander von Humboldt Foundation; the Belgian Federal Science Policy Office; the Fonds pour la Formation \`a la Recherche dans l'Industrie et dans l'Agriculture (FRIA-Belgium); the Agentschap voor Innovatie door Wetenschap en Technologie (IWT-Belgium); the Ministry of Education, Youth and Sports (MEYS) of the Czech Republic; the Council of Science and Industrial Research, India; the HOMING PLUS programme of the Foundation for Polish Science, cofinanced from European Union, Regional Development Fund, the Mobility Plus programme of the Ministry of Science and Higher Education, the National Science Center (Poland), contracts Harmonia 2014/14/M/ST2/00428, Opus 2014/13/B/ST2/02543, 2014/15/B/ST2/03998, and 2015/19/B/ST2/02861, Sonata-bis 2012/07/E/ST2/01406; the National Priorities Research Program by Qatar National Research Fund; the Programa Clar\'in-COFUND del Principado de Asturias; the Thalis and Aristeia programmes cofinanced by EU-ESF and the Greek NSRF; the Rachadapisek Sompot Fund for Postdoctoral Fellowship, Chulalongkorn University and the Chulalongkorn Academic into Its 2nd Century Project Advancement Project (Thailand); and the Welch Foundation, contract C-1845. \end{acknowledgments}

\bibliography{auto_generated}

\cleardoublepage \appendix\section{The CMS Collaboration \label{app:collab}}\begin{sloppypar}\hyphenpenalty=5000\widowpenalty=500\clubpenalty=5000\vskip\cmsinstskip
\textbf{Yerevan Physics Institute,  Yerevan,  Armenia}\\*[0pt]
A.M.~Sirunyan,  A.~Tumasyan
\vskip\cmsinstskip
\textbf{Institut f\"{u}r Hochenergiephysik,  Wien,  Austria}\\*[0pt]
W.~Adam,  F.~Ambrogi,  E.~Asilar,  T.~Bergauer,  J.~Brandstetter,  E.~Brondolin,  M.~Dragicevic,  J.~Er\"{o},  M.~Flechl,  M.~Friedl,  R.~Fr\"{u}hwirth\cmsAuthorMark{1},  V.M.~Ghete,  J.~Grossmann,  J.~Hrubec,  M.~Jeitler\cmsAuthorMark{1},  A.~K\"{o}nig,  N.~Krammer,  I.~Kr\"{a}tschmer,  D.~Liko,  T.~Madlener,  I.~Mikulec,  E.~Pree,  D.~Rabady,  N.~Rad,  H.~Rohringer,  J.~Schieck\cmsAuthorMark{1},  R.~Sch\"{o}fbeck,  M.~Spanring,  D.~Spitzbart,  W.~Waltenberger,  J.~Wittmann,  C.-E.~Wulz\cmsAuthorMark{1},  M.~Zarucki
% ...
\vskip\cmsinstskip
\dag:~Deceased\\
d:~Also at Vienna University of Technology,  Vienna,  Austria\\
d:~Also at State Key Laboratory of Nuclear Physics and Technology;~Peking University,  Beijing,  China\\
d:~Also at Universidade Estadual de Campinas,  Campinas,  Brazil\\
d:~Also at Universidade Federal de Pelotas,  Pelotas,  Brazil\\
d:~Also at Universit\'{e}~Libre de Bruxelles,  Bruxelles,  Belgium\\
d:~Also at Institute for Theoretical and Experimental Physics,  Moscow,  Russia\\
d:~Also at Joint Institute for Nuclear Research,  Dubna,  Russia\\
d:~Also at Suez University,  Suez,  Egypt\\
d:~Now at British University in Egypt,  Cairo,  Egypt\\
d:~Also at Fayoum University,  El-Fayoum,  Egypt\\
d:~Now at Helwan University,  Cairo,  Egypt\\
d:~Also at Universit\'{e}~de Haute Alsace,  Mulhouse,  France\\
d:~Also at Skobeltsyn Institute of Nuclear Physics;~Lomonosov Moscow State University,  Moscow,  Russia\\
d:~Also at Tbilisi State University,  Tbilisi,  Georgia\\
d:~Also at CERN;~European Organization for Nuclear Research,  Geneva,  Switzerland\\
d:~Also at RWTH Aachen University;~III.~Physikalisches Institut A, ~Aachen,  Germany\\
d:~Also at University of Hamburg,  Hamburg,  Germany\\
d:~Also at Brandenburg University of Technology,  Cottbus,  Germany\\
d:~Also at MTA-ELTE Lend\"{u}let CMS Particle and Nuclear Physics Group;~E\"{o}tv\"{o}s Lor\'{a}nd University,  Budapest,  Hungary\\
d:~Also at Institute of Nuclear Research ATOMKI,  Debrecen,  Hungary\\
d:~Also at Institute of Physics;~University of Debrecen,  Debrecen,  Hungary\\
d:~Also at Indian Institute of Technology Bhubaneswar,  Bhubaneswar,  India\\
d:~Also at Institute of Physics,  Bhubaneswar,  India\\
d:~Also at University of Visva-Bharati,  Santiniketan,  India\\
d:~Also at University of Ruhuna,  Matara,  Sri Lanka\\
d:~Also at Isfahan University of Technology,  Isfahan,  Iran\\
d:~Also at Yazd University,  Yazd,  Iran\\
d:~Also at Plasma Physics Research Center;~Science and Research Branch;~Islamic Azad University,  Tehran,  Iran\\
d:~Also at Universit\`{a}~degli Studi di Siena,  Siena,  Italy\\
d:~Also at INFN Sezione di Milano-Bicocca;~Universit\`{a}~di Milano-Bicocca,  Milano,  Italy\\
d:~Also at Purdue University,  West Lafayette,  USA\\
d:~Also at International Islamic University of Malaysia,  Kuala Lumpur,  Malaysia\\
d:~Also at Malaysian Nuclear Agency;~MOSTI,  Kajang,  Malaysia\\
d:~Also at Consejo Nacional de Ciencia y~Tecnolog\'{i}a,  Mexico city,  Mexico\\
d:~Also at Warsaw University of Technology;~Institute of Electronic Systems,  Warsaw,  Poland\\
d:~Also at Institute for Nuclear Research,  Moscow,  Russia\\
d:~Now at National Research Nuclear University~'Moscow Engineering Physics Institute'~(MEPhI), ~Moscow,  Russia\\
d:~Also at St.~Petersburg State Polytechnical University,  St.~Petersburg,  Russia\\
d:~Also at University of Florida,  Gainesville,  USA\\
d:~Also at P.N.~Lebedev Physical Institute,  Moscow,  Russia\\
d:~Also at INFN Sezione di Padova;~Universit\`{a}~di Padova;~Universit\`{a}~di Trento~(Trento), ~Padova,  Italy\\
d:~Also at Budker Institute of Nuclear Physics,  Novosibirsk,  Russia\\
d:~Also at Faculty of Physics;~University of Belgrade,  Belgrade,  Serbia\\
d:~Also at University of Belgrade;~Faculty of Physics and Vinca Institute of Nuclear Sciences,  Belgrade,  Serbia\\
d:~Also at Scuola Normale e~Sezione dell'INFN,  Pisa,  Italy\\
d:~Also at National and Kapodistrian University of Athens,  Athens,  Greece\\
d:~Also at Riga Technical University,  Riga,  Latvia\\
d:~Also at Universit\"{a}t Z\"{u}rich,  Zurich,  Switzerland\\
d:~Also at Stefan Meyer Institute for Subatomic Physics~(SMI), ~Vienna,  Austria\\
d:~Also at Adiyaman University,  Adiyaman,  Turkey\\
d:~Also at Istanbul Aydin University,  Istanbul,  Turkey\\
d:~Also at Mersin University,  Mersin,  Turkey\\
d:~Also at Cag University,  Mersin,  Turkey\\
d:~Also at Piri Reis University,  Istanbul,  Turkey\\
d:~Also at Izmir Institute of Technology,  Izmir,  Turkey\\
d:~Also at Necmettin Erbakan University,  Konya,  Turkey\\
d:~Also at Marmara University,  Istanbul,  Turkey\\
d:~Also at Kafkas University,  Kars,  Turkey\\
d:~Also at Istanbul Bilgi University,  Istanbul,  Turkey\\
d:~Also at Rutherford Appleton Laboratory,  Didcot,  United Kingdom\\
d:~Also at School of Physics and Astronomy;~University of Southampton,  Southampton,  United Kingdom\\
d:~Also at Instituto de Astrof\'{i}sica de Canarias,  La Laguna,  Spain\\
d:~Also at Utah Valley University,  Orem,  USA\\
d:~Also at Beykent University,  Istanbul,  Turkey\\
d:~Also at Bingol University,  Bingol,  Turkey\\
d:~Also at Erzincan University,  Erzincan,  Turkey\\
d:~Also at Sinop University,  Sinop,  Turkey\\
d:~Also at Mimar Sinan University;~Istanbul,  Istanbul,  Turkey\\
d:~Also at Texas A\&M University at Qatar,  Doha,  Qatar\\
d:~Also at Kyungpook National University,  Daegu,  Korea\\
\end{sloppypar}
\end{document}